\documentclass[onecolumn,10pt]{article}

\usepackage[utf8]{inputenc} 

\usepackage[a4paper,margin=2.54cm]{geometry}
\usepackage{flushend}
\usepackage{qcircuit}
\usepackage{caption}
\usepackage{amsmath,amsthm}

\usepackage{bm}  
\usepackage{bbm}
\usepackage{verbatim}
\usepackage{xcolor}

\usepackage[backend=biber,sorting=none,style=phys,biblabel=brackets,backref=false,bibencoding=utf8,maxnames=20,eprint=true]{biblatex}
\DefineBibliographyStrings{english}{%
  backrefpage = {page},
  backrefpages = {pages},
}
\usepackage[bookmarks,colorlinks,breaklinks]{hyperref}  
\definecolor{dullmagenta}{rgb}{0.4,0,0.4}   
\definecolor{darkblue}{rgb}{0,0,0.4}
\hypersetup{linkcolor=red,citecolor=blue,filecolor=dullmagenta,urlcolor=darkblue} 

\usepackage[bitstream-charter]{mathdesign}
\usepackage[T1]{fontenc}

\usepackage{titlesec}
\titleformat*{\section}{\bfseries}
\titleformat*{\subsection}{\normalsize\bfseries}
\titleformat*{\subsubsection}{\bfseries}
\titleformat*{\paragraph}{\large\bfseries}
\titleformat*{\subparagraph}{\large\bfseries}

\titlespacing\section{0pt}{12pt plus 4pt minus 2pt}{2pt plus 2pt minus 2pt}

\usepackage{titling}
\posttitle{\par\end{center}}
\predate{}
\postdate{}
  \DeclareFontFamily{U}{mathb}{\hyphenchar\font45}
\DeclareFontShape{U}{mathb}{m}{n}{
      <5> <6> <7> <8> <9> <10> gen * mathb
      <10.95> mathb10 <12> <14.4> <17.28> <20.74> <24.88> mathb12
      }{}
\DeclareSymbolFont{mathb}{U}{mathb}{m}{n}

\DeclareFontFamily{U}{matha}{\hyphenchar\font45}
\DeclareFontShape{U}{matha}{m}{n}{
      <5> <6> <7> <8> <9> <10> gen * matha
      <10.95> matha10 <12> <14.4> <17.28> <20.74> <24.88> matha12
      }{}
\DeclareSymbolFont{matha}{U}{matha}{m}{n}

\DeclareMathSymbol{\oasterisk}{3}{matha}{"66}
\DeclareMathSymbol{\boxasterisk}{3}{mathb}{"66}

\usepackage{graphicx}

\newcommand{\ket}[1]{\left|#1\right\rangle}

\newcommand{\bra}[1]{\left\langle #1\right|}

\newcommand{\ketbra}[1]{\ket{#1}\bra{#1}}

\newcommand{\id}{\mathbbm 1}

\usepackage{xfrac}

\begin{document}

\author{
{\normalsize Joseph M.~Renes}\\
\emph{\normalsize Institute for Theoretical Physics, ETH Zurich, 8093 Z\"urich, Switzerland}
}

\title{\large {\bf Belief propagation decoding of quantum channels by passing quantum messages}}

\date{\vspace{-\baselineskip}}

\maketitle

\begin{abstract}

The belief propagation algorithm is a powerful tool in a wide range of disciplines from statistical physics to machine learning to computational biology, and is ubiquitous in decoding classical error-correcting codes.
The algorithm works by passing messages between nodes of the factor graph associated with the code and enables efficient decoding of the channel, in some cases even up to the Shannon capacity. 
Here we construct the first belief propagation algorithm which passes \emph{quantum messages} on the factor graph and is capable of decoding the classical-quantum channel with pure state outputs. 
This gives explicit decoding circuits whose number of gates is quadratic in the code length.  
We also show that this decoder can be modified to work with polar codes for the pure state channel and as part of a decoder for transmitting quantum information over the amplitude damping channel. These represent the first explicit capacity-achieving decoders for non-Pauli channels. 
\end{abstract}

\section{Introduction}
Graphical models are at the heart of the current revolution in machine learning and computational statistics. They provide simple representations of the correlations among large numbers of random variables and enable efficient algorithms for feature discovery and analysis. 
Among the most well-known of these algorithms is belief propagation (BP), whose origin can be traced to the Bethe-Peierls approximation in statistical physics~\cite{mezard_information_2009}.
BP can be used to marginalize the joint distribution of several random variables, often efficiently. For instance, in the setting of reliable communication over noisy channels via error correction, BP is used to find the most likely input for a given set of observed outputs. 
Indeed, in modern coding theory BP is simply indispensible~\cite{richardson_modern_2008}. 
The joint distribution of channel inputs and outputs can be represented by a factor graph, and BP works by passing messages between the nodes of this graph (an instance of more general message-passing algorithms). 
This leads to efficient decoding algorithms for high rate codes, several of which are employed in current wireless communication standards. 
Moreover, it was recently shown that belief propagation decoding of a certain class of low-density parity-check (LDPC) codes can achieve the Shannon capacity~\cite{kudekar_spatially_2013}.


Factor graphs have been adapted to the quantum-mechanical setting from several different perspectives~\cite{tucci_quantum_1999,leifer_quantum_2008,loeliger_factor-graph_2012,loeliger_factor_2015}. 
Applied to quantum communication, BP and other message passing methods have been constructed for syndrome decoding of a variety of stabilizer codes subjected to Pauli noise channels~\cite{ollivier_description_2003-1,mackay_sparse-graph_2004,poulin_optimal_2006,poulin_iterative_2008,leifer_quantum_2008,poulin_quantum_2009,duclos-cianci_fast_2010,ferris_tensor_2014}.
Despite their use in decoding quantum codes, these message passing algorithms are classical. 
Indeed, decoding any stabilizer code used for a Pauli channel or the erasure channel is essentially a classical task due to the Gottesman-Knill theorem~\cite{gottesman_heisenberg_1998}. 
However, stabilizer decoding is not optimal for non-Pauli channels such as the amplitude damping channel, for either the entanglement fidelity achievable by fixed-size codes or the largest achievable rates for codes with increasing blocklength. Therefore it would be of interest to extend BP decoding to more general channels. 
As much also holds in the setting of quantum polar codes, where the classical decoding method (ultimately a variant of BP) can only be employed without loss of rate for Pauli channels or the erasure channel~\cite{renes_efficient_2012,wilde_towards_2013,renes_polar_2014}. 

Note that the quantum decoding problem is different than the one solved by the classical algorithm for ``quantum belief propagation'' in~\cite{leifer_quantum_2008}.\footnote{The algorithm of \cite{hastings_quantum_2007} is also a classical algorithm.}  
There, one is interested in computing marginals of quantum states which have a structure given by a factor graph. 
For classical decoding, computing such marginals is indeed sufficient, as we will describe in more detail below. 
But even for bitwise decoding of a classical-quantum (CQ) channel having classical input and quantum output, it is not enough to know the relevant marginal state; we need a way to perform the optimal (Helstrom) measurement~\cite{helstrom_quantum_1976} or some suitable approximation. 
Put differently, a quantum BP decoder is a quantum algorithm, and we may expect that it will need to pass quantum messages.

In this paper we construct a quantum BP decoding algorithm for the pure state channel, a binary input CQ channel whose outputs are pure states. 
The algorithm for estimating a single input bit works by passing single qubits as well as classical information along the factor graph, while sequential estimation of all input bits requires passing many qubits.  
For codes whose factor graphs are trees, as well as for polar codes, we show how the BP decoder leads to explicit circuits for the optimal measurement that have quadratic size in the code length. 
To the best our knowledge, this is the first instance of a quantum algorithm for belief propagation.

The pure state channel arises, for instance, in binary phase-shift keying (BPSK) modulation of a pure loss Bosonic quantum channel, whose channel outputs are coherent states~\cite{guha_polar_2012}. 
Thus, our result gives an explicit construction of a successive cancellation decoder for the capacity-achieving polar code described in \cite{guha_polar_2012}, and addresses the issue of decoding CQ polar codes discussed in \cite{wilde_towards_2013}. 
Moreover, 
the pure state channel also arises as part of the quantum polar decoder for the amplitude damping channel~\cite{renes_efficient_2012,renes_polar_2014}, and therefore our result gives an explicit decoder for polar codes over this channel.

The remainder of the paper is structured as follows. 
In the next section give a very brief overview of factor graphs and their use in classical decoding, and then rewrite the BP rules in a manner that lead to the quantum algorithm. Section~\ref{sec:qBP} gives the quantum BP decoding algorithm and applications to polar codes are given in Section~\ref{sec:polar}. We finish with several open questions for future research raised by our result. 

\section{Belief propagation decoding on factor graphs}
\label{sec:classicalBP}
Let us first examine BP on factor graphs directly in the coding context; for a more general treatment see~\cite{mackay_information_2002,richardson_modern_2008}. 
Consider the problem of reliable communication over a memoryless channel $W$ using a linear code $C$. 
Fix $C$ to be an $n$-bit code, i.e.\ a linear subspace of $\mathbb Z_2^n$, and suppose that the channel $W$ maps inputs in $\mathcal X=\mathbb Z_2$ to some alphabet $\mathcal Y$ according to the transition probabilities $P_{Y|X=x}=W(y|x)$. 
Now suppose a codeword $x_1^n=(x_1,x_2,\dots,x_n)\in C$ is picked at random and its consituent bits are each subjected to $W$, producing the output $y_1^n$. 
The goal of decoding is to invert this process and determine the input codeword from the channel output. 
This is a task of statistical inference, whose nominal solution is to output the $x^n_1$ which maximizes the conditional probability of inputs given outputs, $P_{X^n|Y^n}$. 
Since we assume the inputs are uniformly chosen from $C$, we can directly work with the joint distribution $P_{X^nY^n}$ of inputs and outputs.
In general, though, this task is known to be computationally intractable.

\begin{figure}
\centering
\includegraphics{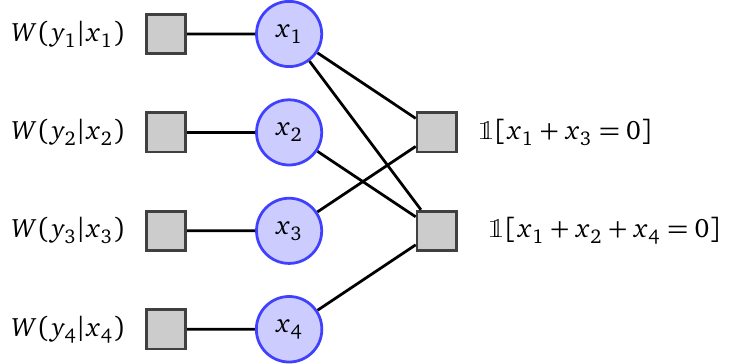}
\caption{\label{fig:fgrepcode} Factor graph for the joint probability distribution of a four-bit code with two parity checks $x_1+x_3=0$ and $x_1+x_2+x_4=0$.}
\end{figure}

A simpler approach is to decode bitwise and find the most likely value of $x_k$ given $y_1^n$, for each $k$. 
Then we are interested in the marginal distribution $P_{X_kY^n}$, and we need only determine which of the two values of $x_k$ maximize $P_{X_kY^n}(x_k,y_1^n)$. 
Exact marginalization is also generally computationally intractable since the size of the joint distribution grows exponentially in the number of variables.
However, for linear codes the joint distribution can be factorized, which often greatly simplifies the marginalization task. 
The joint distribution $P_{X^nY^n}$ can be written 
\begin{align}
\label{eq:jointprob}
P_{X^nY^n}(x_1^n,y_1^n)=\frac1{|C|}\id[{x^n_1\in C}]\prod_{j=1}^n W(y_j|x_j).
\end{align}
Since the channel is memoryless, the channel contribution to \eqref{eq:jointprob} is already in factorized form.
Meanwhile, code membership is enforced by a sequence of parity-check constraints associated with the code, which also leads to factorization.
In the three-bit repetition code, for instance, there are two parity constraints, $x_1+x_2=0$ and $x_2+x_3=0$ (or $x_1+x_3=0$), and therefore $\id[{x_1^3\in C}]=\id[x_1+x_2=0]\,\id[x_2+x_3=0]$. 
We can represent the joint distribution of any linear code (up to normalization) by a factor graph; Figure~\ref{fig:fgrepcode} shows the factor graph of a code involving two parity checks on four bits. 
For an arbitrary factorizeable function, the factor graph contains one (round) variable node for each variable and one (square) factor node for each factor, and factor nodes are connected to all their constituent variable nodes. 
This convention is violated in the figure by not including $y_j$ variable nodes; instead they are treated as part of the channel factors since their values are fixed and in any case each is connected to only one factor node.  

For factor graphs which are trees, meaning only one path connects any two nodes as in  Figure~\ref{fig:fgrepcode}, the belief propagation algorithm can compute the marginal distributions exactly. 
In the present context of coding, it directly finds the most likely input value. 
Supposing we are interested in determining $x_1$, treat variable node $x_1$ as the root of the tree. 
BP then proceeds by passing messages between nodes, starting from the leaves (here, channel outputs) and working inward, combining all relevant information as it goes. 
Simplifying the general BP rules (see \cite{richardson_modern_2008}) to the decoding problem, the initial messages from the channel factors to the variable nodes can be taken as the log-likelihood ratios $\ell=\log [W(y_j|0)/W(y_j|1)]$ of the channel given the output $y_j$ (here we suppress the dependence of $\ell$ on the channel output $y_j$). 
At variable nodes the messages simply add, so that the outgoing $\ell$ is the sum of incoming $\ell_k$. 
At check nodes the rule is more complicated: $\tanh\frac\ell 2 =\prod_k \tanh \frac{\ell_k}2$. 
After all messages have arrived at the root, the algorithm produces the log-likelihood ratio for $x_1$ given all the channel outputs, and the decoder simply outputs 0 if the ratio is positive or 1 if negative. 

By adopting a modified update rule it is in fact possible to compute all the marginals at once with only a modest overhead. 
Instead of only proceeding inward from the leaves, we send messages in both directions along each edge, starting by sending channel log-likelihoods in from the leaves. 
Each node sends messages on each edge once it has received messages on all its other edges.  
For graphs that contain loops, the algorithm is not guaranteed to converge, but one can nevertheless hope that the result is a good approximation and that the decoder outputs the correct value. 
This is borne out in practice for turbo codes and LDPC codes. 

There is an intuitive way of understanding the BP decoding algorithm which is the basis of our quantum generalization. 
At every step the message can be interpreted as the log-likelihood ratio of the effective channel from that node to its descendants. This is sensible as the likelihood ratio is a sufficient statistic for estimating the (binary) input from the channel output. 
The rules for combining messages can then be interpreted as rules for combining channels, and the algorithm can be seen as successively simplifying the channel from the root to the leaves by utilizing the structure of the factor graph.
At variable nodes, adding the log-likelihood ratios for two channels $W$ and $W'$ amounts to considering the convolution channel $W\oasterisk W'$ with transition probabilities given by 
\begin{align}
\label{eq:variablenode}
[W\oasterisk W'](y,y'|x)=W(y|x)W'(y'|x).
\end{align}
That is, the effective channel associated with a variable node is simply the convolution $W_1\oasterisk\cdots \oasterisk W_k$ of its descendants. 
The form of the effective channel at check nodes is not as immediate, but it is not too difficult to verify that the appropriate channel convolution $W\boxasterisk W'$ has transition probabilities
\begin{equation}
\begin{aligned}
\label{eq:checknode}
[&W\boxasterisk W'](y,y'|x)=\tfrac12\left(W(y|x)W'(y'|0)+W(y|x+1)W'(y'|1)\right).
\end{aligned}
\end{equation}

These two channel convolutions are also the fundamental building blocks of polar codes~\cite{arikan_channel_2009}, at least when the input channels are symmetric. 
The check node convolution is the ``worse'' channel in the channel splitting or channel synthesis step (cf.\ \cite[Eq.\ 19]{arikan_channel_2009}); this holds regardless of the symmetry of the channel. 
On the other hand, the ``better'' combination of $W$ and $W'$ is defined by (cf.\ \cite[Eq.\ 20]{arikan_channel_2009}) $W''(y,y',x|x')=\tfrac12W(y|x+x')W'(y'|x')$. 
Compared to \eqref{eq:variablenode}, the input $x$ is uniformly random and not always zero, but it is given at the channel output. 
When $W$ is symmetric in the sense that $W(y|x+u)=W(\pi_u(y)|x)$ for a suitable permutation $\pi$ of the output alphabet depending on $u$, we can reversibly transform $W''$ into $W\oasterisk W'$ and vice versa. 

\section{Belief propagation decoding of quantum outputs} 
\label{sec:qBP}
The form of the check and variable convolutions also applies to channels with quantum output.\footnote{This was first applied in the setting of polar codes in~\cite{wilde_polar_2013}.} 
We need only replace the probability distributions over the output alphabet by quantum states. 
Abusing notation, let us denote by $W(x)$ the quantum state of the output of $W$ given input $x$. 
This includes the previous case by considering commuting $W(x)$. 
The the variable and check node convolutions are now just
\begin{align}
[W\oasterisk W'](x)&=W(x)\otimes W'(x),\qquad \text{and}\\
[W\boxasterisk W'](x) &= \tfrac12(W(x)\otimes W'(0)+W(x{+}1)\otimes W'(1)).
\end{align}

To properly generalize the BP decoding algorithm we need a ``sufficient statistic'' for the quantum channels at the various nodes. 
For binary-input pure state channels, it turns out that a combination of classical bits and just one qubit suffices. 
The channel outputs can always be represented by a qubit, so suppose that $W$ outputs $\ket{\pm \theta}$, where $\ket{\theta}=\cos\frac\theta 2\ket{0}+\sin\frac\theta 2\ket 1$. 
Note that the overlap of the two states is $\cos\theta$ and the Helstrom measurement for these two states is measurement of the $\sigma_x$ operator. 

The convolution $W\oasterisk W'$ outputs either $\ket{ \theta}\otimes \ket{ \theta'}$ or $\ket{- \theta}\otimes \ket{- \theta'}$, which are again two pure states, with an overlap angle $\theta^\oasterisk$ given by $\cos\theta^\oasterisk=\cos\theta\cos\theta'$. 
The following unitary transformation compresses the states to the first qubit, leaving the second in the state $\ket{0}$:
\begin{align}
U_\oasterisk(\theta,\theta')=
\begin{pmatrix} 
a_+ & 0 & 0 & a_-\\ 
a_- & 0 & 0 & -a_+\\
0 & b_+ & b_- & 0\\
0 & b_- & -b_+ & 0
\end{pmatrix},
\end{align}
with $a_\pm\sqrt{1+\cos\theta\cos\theta'}=\tfrac1{\sqrt2}(\cos(\frac{\theta-\theta'}2)\pm\cos(\frac{\theta+\theta'}2))$ and 
$b_\pm\sqrt{1-\cos\theta\cos\theta'}=\tfrac1{\sqrt2}(\sin(\frac{\theta+\theta'}2)\mp\sin(\frac{\theta-\theta'}2))$.
To combine more than two channels, we just perform the pairwise convolution sequentially. 
Thus, the $\oasterisk$ convolution of pure state channels can itself be represented as a pure state channel.

The $\boxasterisk$ convolution is more complicated because the outputs are no longer pure. 
However, applying the unitary $U_\boxasterisk=\textsc{cnot}_{1\to 2}$ results in a CQ state of the form $\sum_{j\in\{0,1\}} p_j\ketbra{\pm \theta^\boxasterisk_j}\otimes \ketbra j$.
We are free to measure the second qubit, and conditional state of the first qubit is again one of two pure states, though now the overlap $\cos\theta^\boxasterisk_j$ depends on the measurement outcome $j$. 
In particular, $p_0=\tfrac12(1+\cos\theta\cos\theta')$, $p_1=1-p_0$, and the two overlaps are given by 
\begin{align}
\label{eq:box1}
\cos\theta^\boxasterisk_0 &=\frac{\cos\theta+\cos\theta'}{1+\cos\theta\cos\theta'},\\
\cos\theta^\boxasterisk_1 &=\frac{\cos\theta-\cos\theta'}{1-\cos\theta\cos\theta'}.
\label{eq:box2}
\end{align}
For outcome $j=0$ the angle between the states has decreased, while for outcome $j=1$ the angle has increased.
Therefore, the $\boxasterisk$ convolution of pure state channels can be represented by two pure state channels, corresponding to the two measurement outcomes. As before, several channels can be combined sequentially.


The quantum decoding algorithm now proceeds as in classical BP, taking the quantum outputs of the channels and combining them at variable and check nodes. 
At a variable node the algorithm combines the outputs using $U_\oasterisk$ and forwards the output to its parent node. At check nodes the algorithm applies $U_\boxasterisk$, measures the second qubit, and forwards both the qubit and the measurement result to its parent node. The classical messages are required to inform parent variable nodes how to choose the angles in subsequent $U_\oasterisk$ unitaries. 
Ultimately this procedure results in one qubit at the root node such that measurement of $\sigma_x$ corresponds to the optimal Helstrom measurement for the associated bit. This then is sufficient to estimate one input bit. 

For example, return to the code depicted in Figure~\ref{fig:fgrepcode} for a pure state channel with overlap $\theta$, and suppose we are interested in decoding the first bit. 
Starting at the leaves, the outputs of all but the first channel can be immediately passed to their corresponding variable nodes, since these variable nodes do not have any other outward branches. (Formally this follows from the convolution rules by considering convolution with a trivial channel, having $\theta=0$.) 
The output of the first channel, meanwhile, must wait to be combined according to the $\oasterisk$ convolution with several other qubit messages. 
Next, since 2 and 4 are connected by a check node, we combine qubits 2 and 4 into one qubit (2) and one classical bit (4) by applying $U_\boxasterisk$ and measuring the 4th qubit. 
As qubits 1 and 3 are connected by a variable node, we can simultaneously combine these with $U_\oasterisk(\theta,\theta)$. 
Finally, we combine qubits 1 and 2 by applying $U_\oasterisk(\theta^\oasterisk,\theta^\boxasterisk_j)$, where $\cos\theta^\oasterisk=\cos^2\theta$ and $\cos\theta_{0}^\boxasterisk=\frac{2\cos\theta}{(1+\cos^2\theta)}$, $\theta_1^\boxasterisk=\frac\pi 2$, depending on the value $j$ of the earlier measurement. 
A quantum circuit implementing these steps is shown in Figure~\ref{fig:circuit}.

\begin{figure}
\centering
\[\Qcircuit @C=1.5em @R=.7em {
   & \qw         & \multigate{1}{U_\oasterisk}   & \qswap      & \qw{|}\\
   & \qswap     & \ghost{U_\oasterisk}          & \qswap \qwx & \multigate{1}{U_\oasterisk} & \gate{H}&\meter & \cw\\
   & \qswap \qwx  & \multigate{1}{U_\boxasterisk} & \qw         & \ghost{U_\oasterisk} & \qw{|}\\
   & \qw       & \ghost{U_\boxasterisk}        & \qw         & \meter  \cwx  & 
}
\]
\caption{\label{fig:circuit} Circuit decoding the first bit of the code depicted in Figure~\ref{fig:fgrepcode}. 
The first $\oasterisk$ convolution is $U_\oasterisk(\theta,\theta)$, the second $U_\oasterisk(\theta^\oasterisk,\theta^\boxasterisk_j)$ for $\cos\theta^\oasterisk=\cos^2\theta$ and $\cos\theta_{0}^\boxasterisk=\frac{2\cos\theta}{(1+\cos^2\theta)}$, $\theta_1^\boxasterisk=\frac\pi 2$, depending on the value $j$ of the measurement outcome in the bottom wire. The symbol $\dashv$ denotes that the qubit is discarded. The final Hadamard gate and measurement implement the Helstrom measurement.
}
\end{figure}

One drawback is that the above procedure implements the Helstrom measurement destructively, since once we estimate the first bit we no longer have the original channel output in order to estimate the second bit. 
And we cannot run the algorithm backwards to reproduce the channel output as we have made measurements at every check node. 
To implement the Helstrom measurement as nondestructively as possible, we can leave the CQ output states unmeasured and instead use the classical subsystems to coherently control the variable node unitaries $U_\oasterisk$.
In this way the steps in the algorithm can be reversed, save the final measurement. 
For example, in Figure~\ref{fig:circuit} all output qubits are kept and the classical measurement and subsequent conditioning of the second $U_\oasterisk$ gate is performed by a coherent conditional gate involving three qubits. 

Denoting the unitary action of the algorithm for the $j$th bit by $V_j$, the Helstrom measurement can be implemented by the projective measurement with projectors $\Pi_{j,k}=V_j^* |\tilde k\rangle\langle \tilde k|_j V_j$, where $|\tilde k\rangle\langle \tilde k|_j$ denotes the $k$th $\sigma_x$ basis projector on the $j$th qubit. 
Each $V_j$ is composed of $O(n)$ gates, yielding an overall circuit size of $O(n^2)$ to decode all bits.  
Supposing that the code is designed such that the $j$th input bit can be estimated with error no larger than $\epsilon_j$, Gao's non-commutative union bound~\cite{gao_quantum_2015} implies that the error in sequentially estimating all bits is no worse than $4\sum_j \epsilon_j$. 

\section{Applications to polar codes}
\subsection{Polar codes for the pure state channel}
\label{sec:polar}
Polar codes for the pure state channel may also be decoded with this algorithm. 
Indeed, the successive cancellation decoding algorithm proposed by Ar\i{}kan in~\cite{arikan_channel_2009} proceeds precisely by combining channels using the $\oasterisk$ and $\boxasterisk$ rules, and was adapted to the case of classical-quantum channels in~\cite{wilde_polar_2013}. The difference is that successive cancellation does not use the factor graph of the code, but a graph related to a fixed reversible encoding circuit.  
Importantly, the graph associated to each input of the encoding circuit is a tree. 
In fact, each such graph has logarithmic depth from all channel factors to each variable, and every node has degree three.
Unlike the BP decoder, however, the successive cancellation decoder used by polar codes takes previously decoded bits into account. 
But these bits can be handled by the BP decoder since the pure state channel is symmetric in the manner described at the end of \S\ref{sec:classicalBP}. 
There, the value of the previous bits is incorporated into the better channel by  appropriately permuting the output symbols, which is equivalent to flipping the input value. Similarly, for the pure state channel, applying $\sigma_z$ to the output is equivalent to flipping the input.  
Therefore, the quantum BP decoding algorithm gives a successive cancellation decoder for polar codes over the pure loss Bosonic channel using the BPSK constellation~\cite{guha_polar_2012}. 

\subsection{Quantum polar codes for amplitude damping}
The idea behind the quantum polar coding scheme of \cite{renes_efficient_2012,renes_polar_2014} is to decompose the problem of transmitting quantum information over a channel $\mathcal N_{A\to B}$ into transmitting classical information about two conjugate observables, ``amplitude'' and ``phase'', consider polar codes for each subproblem, and then combine the coding schemes using CSS codes at the encoder and coherent sequential decoding of amplitude and phase at the decoder. 
This decoding strategy is depicted in \cite[Fig.\ 3]{renes_efficient_2012} for Pauli channels and \cite[Fig.\ 1]{renes_uncertainty_2016-1} for the general case. 
As detailed in \cite{renes_polar_2014}, the two classical transmission tasks are to transmit ``amplitude'' information over the CQ channel given by $z\to \rho_z=\mathcal{N}(\ketbra z)$ and ``phase'' information over the CQ channel given by $x\to \varphi_x=(Z^x\otimes \id) (\mathcal I\otimes \mathcal N)[\Phi](Z^x\otimes \id)$. 
Here $\ket{z}$ is an arbitrary basis, and we choose that of $\sigma_z$ for convenience, while $\ket{\Phi}_{A'A}=\sum_z\sqrt{p_z}\ket{z}\ket{z}$ is a bipartite pure state in this same basis with coefficients of our choosing. (See \cite{renes_polar_2014} for the precise relation to the conjugate observables $\sigma_x$ and $\sigma_z$.)  

Let us now show how to build a decoder for the amplitude damping channel $\mathcal N_\gamma$ with damping parameter $\gamma\in [0,1]$. 
First note that the amplitude outputs all commute due to the form of $\mathcal N_\gamma$; the amplitude channel is effectively a classical Z channel in which the input 0 is always transmitted perfectly, but the input 1 may decay to 0 with probability $\gamma$.
Therefore we can use the classical polar encoder and decoder for this channel~\cite{honda_polar_2013}. 
Since the Z channel is not symmetric, the optimal input distribution in the capacity formula is not the uniform distribution, but one with probabilities $p$ and $1-p$. 

Now suppose that the bipartite pure state in the phase channel is the state $\ket{\Phi}=\sqrt{p}\ket{00}+\sqrt{1-p}\ket{11}$. 
Abusing notation slightly and denoting the channel outputs $\varphi_\pm$, it is not difficult to verify that for $U=\textsc{cnot}_{A'\to B}$,
\begin{align}
U\varphi_\pm U^*&=(1-\gamma(1-p))\ketbra{\pm\theta_0}\otimes \ketbra{0}+\gamma(1-p)\ketbra 1\otimes \ketbra{1},\qquad \text{with}\\
\cos\theta_0&=\frac{1-2p-\gamma(1-p)}{1-\gamma(1-p)}.
\end{align}
Each of these states is a CQ state with the first qubit pure and the second qubit classical, just as in a $\boxasterisk$ output. 
Given the second qubit, the first is either in the pure state $\ket{\pm \theta_0}$ corresponding to the channel input $\pm$, or the state $\ket{1}$ independently of the input; the latter is equivalent to $\ket{\theta_1=0}$. 
Hence the decoder can begin just as at a $\boxasterisk$ step, measuring the second qubit to determine the angle associated to the first qubit. 

The rate achievable by the quantum polar code construction is simply $R=\max_{p\in [0,1]} \left(1-H(Z|B)_\psi-H(X|BA')_\xi\right)$, where $\psi_{ZB}=p\ketbra 0\otimes \rho_0+(1-p)\ketbra 1\otimes \rho_1$ and $\xi_{XBA'}=\tfrac12\sum_{x\in \{0,1\}}\ketbra x\otimes \varphi_x$.  
A cumbersome but straightforward calculation confirms that $R$ equals the capacity of the channel, $C(\mathcal N_\gamma)=\max_{p\in[0,1]} \left(h_2((1-\gamma)p)-h_2(\gamma p)\right)$, for $h_2$ is the binary entropy~\cite[Prop.\ 24.7.2]{wilde_quantum_2017}.
Moreover, since the amplitude damping channel is degradable, the arguments in \cite{renes_efficient_2012} ensure that no entanglement-assistance is required to meet the CSS constraint when constructing the quantum polar code.

\section{Discussion}
We have presented a belief propagation algorithm for bitwise decoding of CQ channels which operates by passing quantum messages on tree factor graphs, and shown several applications to polar codes. 
This invites the study of quantum message passing algorithms, and not just in the context of decoding. More generally we may look for BP and related algorithms for any task of statistical inference where the input data comes in the form of many quantum bits, for instance in quantum metrology.  
This work also raises many interesting questions. 
Most immediately in the context of decoding is whether the complexity of the algorithm can be reduced for structured factor graphs. 
Classical polar codes, for instance, have decoding complexity $O(n\log n)$. 
Can this also be achieved for the pure state channel? Similarly, can one find a quantum version of the max-product or Viterbi algorithm for determining the most likely $x_1^n$ given the channel outputs? 

More generally, it would be very interesting to understand how to run the algorithm on a factor graph with loops, or how it can be modified to handle some set of non-pure output states. In the former case it may be useful to explore the characterization of loopy BP as a variational problem~\cite{wainwright_graphical_2007,mezard_information_2009}.
Perhaps in the latter case one can make use of the work on quantum sufficiency  (see e.g.\ \cite{jencova_sufficiency_2006,buscemi_comparison_2012} and references therein) to find a suitable set of quantum messages for a given decoding problem. 

Another interesting question with potentially far-reaching consequences is the relation of the BP algorithm to tensor network methods. 
The problem of marginalization in the commutative setting is explicitly treated as tensor network contraction in~\cite{ferris_tensor_2014}, and the particulars of the quantum BP decoder bear a similarity with the data gathering approach using tensor network states in \cite{blume-kohout_quantum_2013}.
Can the methods of approximating quantum states by tensor networks be used to create efficient approximate decoders? \\[-2mm]



\section*{Acknowledgments} 
It is a pleasure to acknowledge helpful conversations with R\"udiger Urbanke, Marco Mondelli, and David Sutter. 
Thanks also to Narayanan Rengaswamy for pointing out an error in $U_\boxasterisk$ in a previous version of this paper.
This work was supported by the Swiss National Science Foundation (SNSF) via the National Centre of Competence in Research ``QSIT'', and by the European Commission via the project ``RAQUEL''.

\printbibliography[heading=bibintoc,title=References]

\end{document}